\documentclass[prl,aps,twocolumn,amsfonts,amssymb,amsmath,floatfix,showpacs,superscriptaddress,nobibnotes]{revtex4}
\usepackage{graphicx,graphics}
\usepackage{colortbl}
\usepackage{dcolumn}
\usepackage{bm}
\begin{document} 
\title{
Short Range Smectic and Long Range Nematic Order in the Pseudogap
Phase of Cuprates
}
\author{R.S. Markiewicz}
\affiliation{ Physics Department, Northeastern University, Boston MA
02115, USA}
\author{J. Lorenzana} 
\affiliation{ISC-CNR, Dipartimento di Fisica, Universit\`a ``La Sapienza'',
Piazzale Aldo Moro 5, Roma, Italy} 
\author{G. Seibold} 
\affiliation{Institut F\"ur Physik, BTU Cottbus, PBox 101344, 03013 Cottbus,
Germany}   
\author{A. Bansil}
\affiliation{ Physics Department, Northeastern University, Boston MA
02115, USA}
\begin{abstract}
We present a model for the combined nematic and `smectic' or stripe-like orders seen in recent scanning tunneling microscopy (STM) experiments in cuprates.  We model the stripe order as an electronic charge density wave with associated Peierls distortion -- a `Pomeranchuk wave'.  Disorder restricts this primary order to nanoscale domains, while secondary coupling to strain generates nematic order with considerably longer range.
\end{abstract} 
\maketitle

Two of the most intriguing phases in strongly correlated matter are the 
Mott and high-$T_c$ superconducting phases found in close proximity in cuprates.  Any direct connection between these phases is blocked by the universal intervention of a pseudogap phase in the underdoped regime. Controversy swirls about this phase - is it associated with preformed pairs or a competing order, and if the latter, what is the nature of this `hidden order'?  While there is considerable evidence for `smectic' or stripe-like order in underdoped cuprates\cite{Rev1,Rev2}, 
such stripes should be sensitive to the quenched disorder that is generally present in cuprates.  Hence, long range stripe order is less likely than a frozen, glassy phase with only short-range orientational order.  Alternatively, there could be a fluctuating stripe phase consistent with a nematic order, which retains orientational symmetry-breaking while translational symmetry is restored.  Totally unexpected therefore was the recent discovery in Bi$_2$Sr$_2$CaCu$_2$O$_8$ (Bi2212) of a coexisting  short range (glassy) smectic order and a long range nematic order\cite{Seamus}.  In Ref.~\onlinecite{Mesaros}, it was shown that fluctuations of the nematic phase are strongly correlated with the smectic phase, and this coupling was described in terms of a Ginzburg-Landau theory originally developed for liquid crystal phases. 

\begin{figure}
\includegraphics[width=8cm,clip=true]{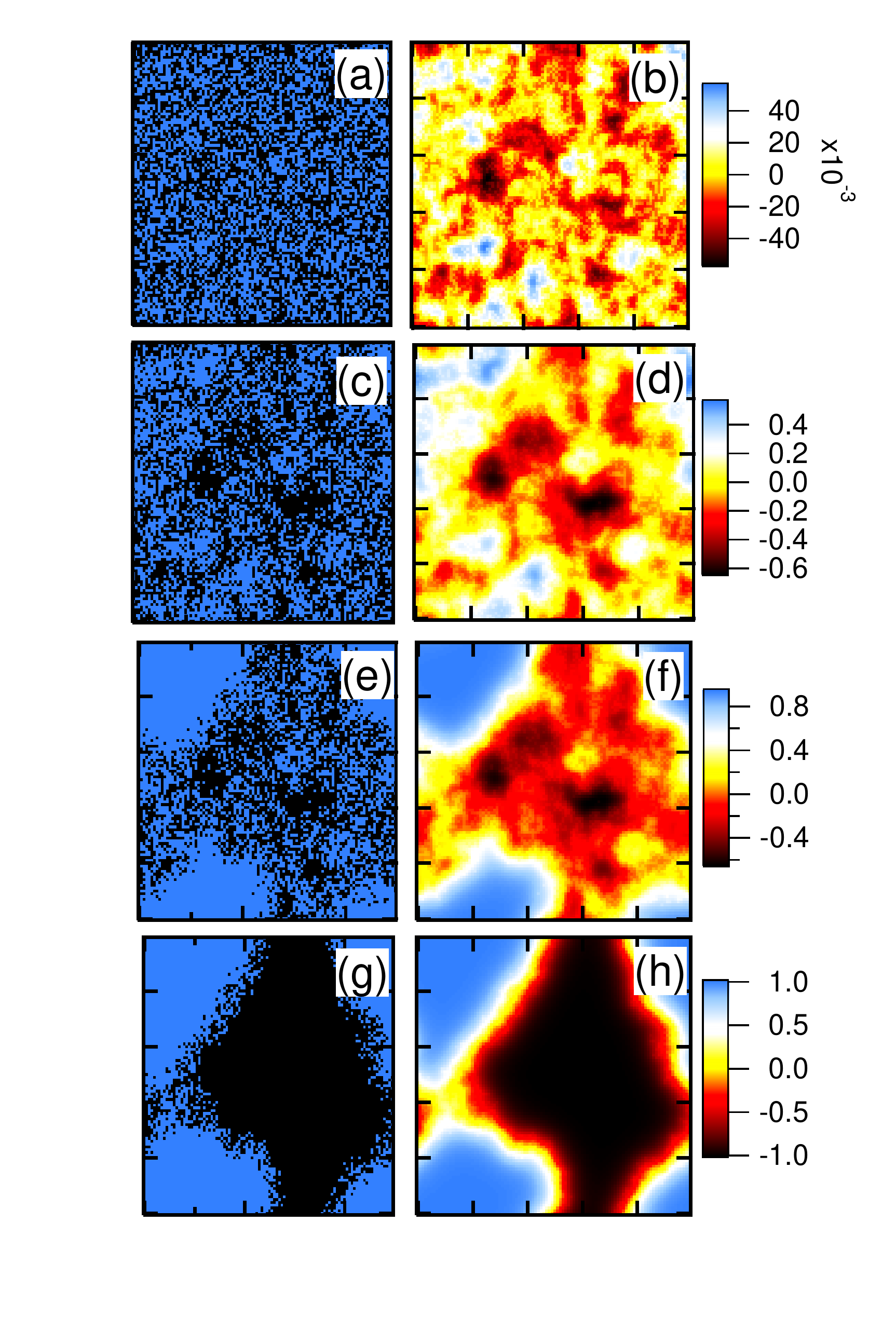}
\caption{(Color online.)
{\bf Co-evolution of DW and shear [or nematic] orders in Bi2201.}
Left side = maps of 100$\times$100 DW domains. with black dots corresponding to $x$-directed stripes ($\sigma_i=+1$) and blue dots to $y$-directed stripes ($\sigma_i=-1$). Right side = corresponding strain fields $\eta$, with magnitude given by color bars.  The different rows correspond to $\delta$ = 0.2 (a,b), 
0.94 (c,d), 0.945 (e,f), and 1.0 (g,h).  Calculations assume periodic boundary conditions, which may limit domain size in frames e-h.Add system dimensions to caption and $\sigma$, $\eta$
to the colorbox.
} 
\label{fig:2} 
\end{figure}

However, the origin of the nematic phase itself was not explained.  In this Letter we provide an explanation for this smectic-nematic phase in terms of a charge-density wave (CDW) coupled to strain.  
We assume that the pseudogap is associated with a competing density wave (DW) order, and that this order has been observed by scanning tunneling microscopy (STM) studies\cite{Yaz,check,HanD,Koh,Wise,JCD,ParkYaz,kos12} of cuprates.  
Then the problem is tackled in two steps. First, we analyze the
occurence of charge order (CO) from the interplay between electrons and lattice 
deformations within a microscopic model. In a second, more phenomenological
approach, we incoporate the effect of impurities which break up the regular  
CO into small domains but still allow for large scale nematic distortions
in agreement with Ref. \cite{Seamus} (cf. Fig. \ref{fig:2}).


In mean-field type approximations of Hubbard models the dominant stripe order appears to be predominantly magnetic\cite{LoSe}.
On the other hand exact diagonalization of small clusters within the tJ-model supplemented with electron-phonon coupling supports the view of stripes which are stabilized by lattice modes \cite{dobriera}.  In fact, experimentally DW order is found to be accompanied by lattice distortions, at least in La$_{2-x}$Sr$_x$CuO$_{4+\delta}$ (LSCO).  Hence we here explore a model with a primary charge order, 
stabilized by a lattice distortion.  Our considerations are based on the time-dependent 
Gutzwiller approximation applied to the one-band Hubbard model
where the electron-lattice coupling is incorporated via a modulation of the hopping parameters.
The calculations are summarized in the supplementary materials. 
Our results are generic for any model of charge-ordered stripes coupled to a lattice distortion -- in particular, the main conclusions should also be valid for magnetically originated stripes with induced charge order or any model of charge-ordered stripes coupled to strain.  
The resulting stripes provide a good description of many 
properties of the modulated phase\cite{Yaz,check,HanD,Koh,Wise,JCD,ParkYaz} found in STM studies of Ca$_2$CuO$_2$Cl$_2$ (CCOC), Bi2212, and Bi$_2$Sr$_2$CuO$_6$ (Bi2201).  



\begin{figure}
\includegraphics[width=10cm,clip=true]{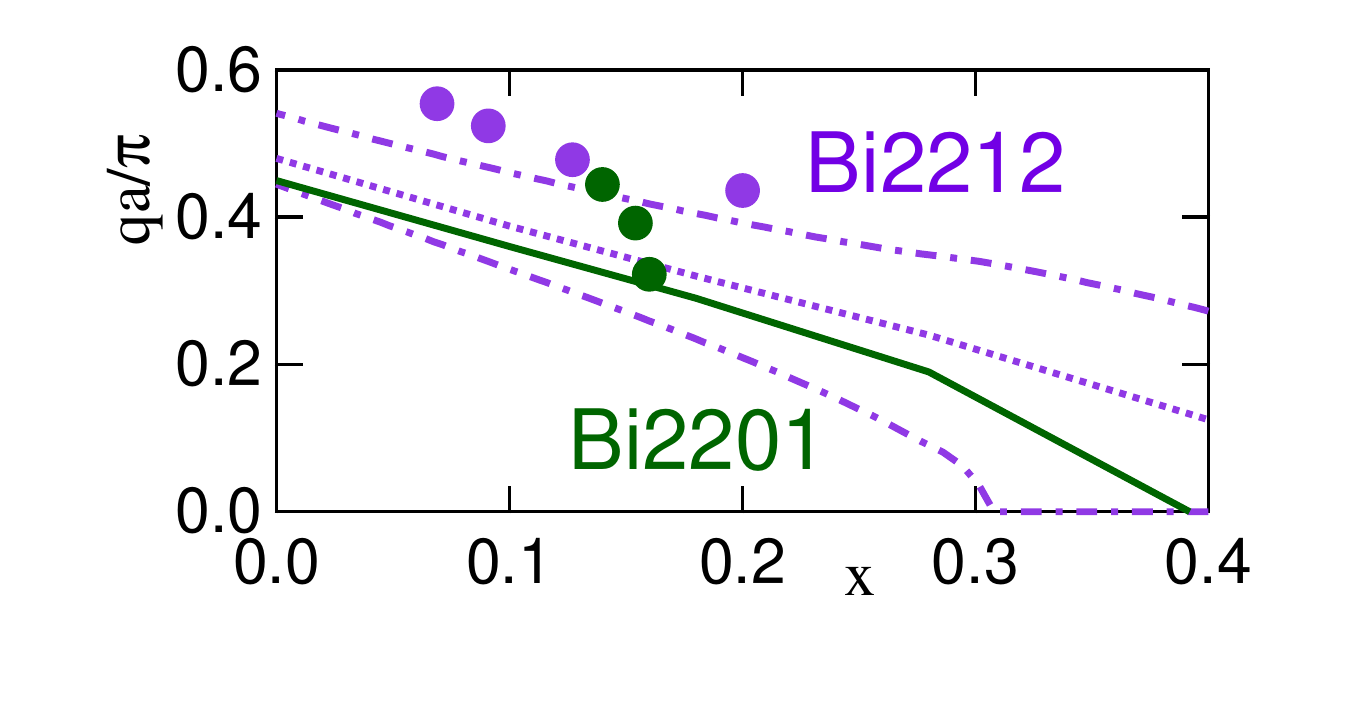}
\caption{(Color online.)
Nesting vectors for ANN $(q,q)$  
charge order, for Bi2201 (green solid line), Bi2212 with (violet dot-dashed) or without (violet dotted) 
bilayer splitting, compared to experimental $(0,q)$ charge stripe vectors for 
Bi2201 (dark green circles)\protect\cite{Wise} and Bi2212 (violet  
circles)\protect\cite{JCD}.  Band parameters for the theoretical calculations are from 
Refs.~\protect\onlinecite{ZX01} and ~\protect\onlinecite{Arun3} respectively.  
} 
\label{fig:1} 
\end{figure}

We find that the stripes are stabilized by nesting of the flat, 
nearly parallel antinodal (AN) FS sections near $(\pi ,0)$ in the Brillouin 
zone,\cite{Kyle,Wise,JCD,ParkYaz} and hence we refer to them as AN nesting (ANN) stripes.  
This ANN phase breaks $C_4$ symmetry and hence also represents a quasi-one-dimensional 
charge order (CO) state.\cite{Koh} Figure~\ref{fig:1} shows that the model correctly 
describes the measured\cite{Wise,JCD} doping dependence of the ANN periodicity.
Figure~\ref{fig:1} plots the calculated doping dependence of the ANN charge nesting 
vectors for the Bi cuprates,\cite{ZX01,Arun3} displaying a strong material 
dependence.  Shown also are the experimental superlattice periodicities for charge order in 
Bi2201\cite{Wise} (green circles) and -2212\cite{JCD} (violet circles). 
Clearly the experimental superlattices in the Bi-cuprates are close to the predicted ANN 
periodicities. However, it should be noted that experimental
data are extracted at the energy scale of the pseudogap whereas the nesting
vectors signal the transition towards CO at zero frequency. This might be 
the reason why the calculated ANN occurs along the diagonals in contrast
to the electronic order along the Cu-O bond direction oberved in STM. 
In Bi2212 there are two nesting vectors, associated with bonding and 
antibonding combinations of the bilayer-split bands, and the experimental data fall close to 
the bonding band nesting vector.  In deeply-underdoped Bi2212, this charge density wave (CDW) may be unstable against nanoscale 
phase separation.\cite{GML}  Hints for this latter effect may have been observed 
in recent STM studies: Ref.~\onlinecite{ParkYaz} found that the phase we here identify as ANN seems to weaken below 1/8th doping, while in a similar doping range in CCOC Ref.~\onlinecite{kos12} found that islands of a phase without $C_4$ symmetry breaking became more prevalent with reduced doping.\cite{neut,eno12}

Allowing for the symmetry-broken CO phase it is found that 
the model can naturally explain a number of features of the pseudogap regime seen in STM and ARPES experiments. This is exemplified for an ordered stripe
array (periodicity $6$ lattice constants) where we refer to the Supplementary
material for details of the corresponding analysis. In particular, we find that (1) there is a {\it contrast reversal} of the stripes in STM images, between positive and negative energies, consistent with experiment\cite{Koh,JCD,Seamus,Madhavan}, Figs.~S3(a,b); (2) maps of the spectral intensity at the Fermi energy typically show arc-like Fermi surfaces (FSs), consistent with ARPES results, which lose intensity close to the conventional $(\pi ,\pi )$-antiferromagnetic (AF) zone boundary as seen in STM\cite{JCD}, Figs.~S3(f,g); (3) by including a secondary magnetic order parameter, it is possible to get strongly angular FS sections, which had been seen in ARPES and taken as evidence for underlying stripe phases\cite{ZaBo}, Fig.~S3(g), or even {\it half-pockets}, which close on the same side of the AF zone boundary, Fig.~S3(e), as seen in ARPES studies\cite{Meso,halfpock1,halfpock2,halfpock3}, although $(\pi /2,\pi /2)$
centered pockets may have a structural origin associated with low-temperature 
tetragonal order\cite{HeShen}, unrelated to stripe physics; (4) ARPES finds that the pseudogap in the antinodal nesting regime is centered at the Fermi level\cite{PJ}, and this arises naturally in the ANN model, since it is precisely the AN FS sections that control the nesting, Fig.~S4.


However, the most puzzling feature of the pseudogap phase is the coexistence of a stripe-like order with a $q=0$ order, which breaks the symmetry of a unit cell -- most likely by making the oxygens inequivalent.\cite{Seamus,Sidis}  Such an extra symmetry breaking arises naturally in any vertical stripes, which have no bond length modulation along $y$ (the direction along the stripe).  Since the hopping parameter $t$ scales as $a^{-p}$, where $a$ is the lattice constant and $p\sim$~3.5 for nearest neighbors, then when $a$ is modulated by $\delta a$ the 
linear corrections average out over the stripes, but there is a {\it finite quadratic correction}, $\delta t_x\sim (\delta a)^2$.  Hence the Pomeranchuk wave couples quadratically to a $q=0$ shear mode, 
$t_x>t_y$.\cite{footx}  This in turn leads to a {\it nematic electronic order}: the hole doping is inequivalent on the in-plane $x$ and $y$ oxygens.\cite{footV,LTT}  
This doping asymetry was recently found\cite{Seamus} in STM studies of Bi2212 and is also 
consistent with the $q=0$ magnetic order found in the pseudogap state of cuprates,\cite{Sidis} 
as long as the oxygen holes have a magnetic moment.  Furthermore, if the stripes fluctuate in 
time or space (due to impurity pinning), then $\delta a=\delta a_0cos(q_xx)$ will average to 
zero, but since $\delta a^2=\delta a_0^2[cos(2q_xx)+1]/2$ the shear contribution at $q=0$ will not.  

Within our model, nematic order is a true {\it emergent} phenomenon.  The primary instability is to a CDW order, but in two dimensions this is unstable in the presence of charged impurities\cite{ImMa}.  On the other hand, there is a secondary coupling to a strain distortion, and this being long-range is much less sensitive to impurities.\cite{BeRL}  Thus stripe domains will be disordered with $x$ and $y$ domains, while shear domains will 
be more robust, having long-range interactions, so the nematic domains will have larger 
correlation lengths, as seen in experiment.  Here we provide a `proof of principle' that the secondary order parameter can have a larger correlation length.  The most general Landau-Ginzburg effective Hamiltonian to model the
strain-density wave interaction is:
\begin{equation}
H_{eff}=H_{\phi}+H_{\eta}+H_{\eta -\phi},
\label{eq:31}
\end{equation}
where the DW order is characterized by a charge density
modulation as\cite{RK2F2}
\begin{equation}
\rho ({\bf r})=\bar\rho +[\phi_x({\bf r})e^{iQ_xx}+\phi_y({\bf r})e^{iQ_yy}
+c.c],
\label{eq:32}
\end{equation}
and $\phi_x$, $\phi_y$ are the two competing density waves in orthogonal
directions.  Then
\begin{eqnarray}
H_{\phi}={\alpha\over 2}\phi^2+{u\over 4}\phi^4+\gamma |\phi_x\phi_y|^2+
{\kappa_L\over 2}[|\partial_x\phi_x|^2+|\partial_y\phi_y|^2]+
\nonumber\\
{\kappa_T\over 2}[|\partial_x\phi_y|^2+|\partial_y\phi_x|^2].
\label{eq:33}
\end{eqnarray}
The strain Hamiltonian is\cite{tweed,tweed2}
\begin{equation}
H_{\eta}={a\over 2}\eta^2+{a_1\over 2}e_1^2+{a_2\over 2}e_2^2
+{\kappa_1\over 2}|{\bf\nabla}\eta |^2,
\label{eq:34}
\end{equation}
with DW-strain coupling
\begin{equation}
H_{\eta -\phi}=\delta\eta [\phi_x^2-\phi_y^2],
\label{eq:35}
\end{equation}
where the strain components are $\eta=(e_{xx}-e_{yy})/\sqrt{2}$, $e_1=(e_{xx}+e_{yy})/\sqrt{2}$, $e_2=e_{xy}$.  
[Since this is a one-band model, the effect of the strain on the oxygens, leading to nematic order, is implicit.]

To compare with STM results, we can simplify the above.  The density wave is 
assumed to be ordered ($\alpha <0$) but strongly pinned, so that it can be 
reduced to an Ising variable,\cite{LCD} $\sigma$ [= + for $\phi_x$, - for 
$\phi_y$].  Then $H_{\phi}$ becomes
\begin{equation}
H_{\phi}=-J\sum_{<ij>}\sigma_i\sigma_j-\sum_ih_i\sigma_i, 
\label{eq:36}
\end{equation}
where each $\sigma_i$ represents the average on a patch, the first sum is over 
nearest neighbors ($<ij>$) on a square lattice of patches, and $h_i$ is a random 
variable in the range $(-h_0,h_0)$.  The strain remains a continuous variable, but now defined on patches.  
Since only the deviatoric strain $\eta$ couples to the stripes, the bulk dilational ($e_1)$ and shear $(e_2)$ strains can be eliminated from the problem.  However, there is a compatibility condition relating the strain components, to satisfy St. Venant's principle, which leads to a long-range interaction of the $\eta$ strains,\cite{tweed2,tweed3}
\begin{equation}
H_{\eta}+H_{\eta -\phi}
=\sum_i[{a\over 2}\eta_i^2+\delta\eta_i\sigma_i+\sum_j{f\cos{[4(\theta_{ij})]}\over (r_i-r_j)^2}(\eta_i-\eta_j)^2],
\label{eq:37}
\end{equation}
where we have used the fact that $\sigma_i$ is an Ising variable, used a renormalized $\delta$, and $\theta_{ij}$ is the angle between grains $i$ and $j$ measured from the Cu-O bond direction.  Numerical results for the DW $\sigma_i$ and strain $\eta_i$ fields are plotted in Fig.~\ref{fig:2}, where we have taken (in units where $a=1$) $h_0=0.7$, $f/D^2=8$ [$D$ is the average patch size], $J=0$ [since the results are not sensitive to small values of this parameter], and let $\delta$ vary.  The small value of $a$ suggests that the strain field is nearly unstable.  This would be expected, since this strain couples strongly to splitting of the VHS peak.

Figure~\ref{fig:2} shows how the random patches of DW order generate strain fields, and how 
the strains develop long-range coherence as the DW-strain coupling is increased.  Frames (a), 
(c), (e), and (g) show the DW domains forfour increasing values of $\delta$, while frames (b), 
(d), (f), and (h) show the corresponding strain fields.  As $\delta\rightarrow 0$, the strain vanishes, 
but even at the low coupling $\delta =0.2$ in Fig.~\ref{fig:2}(b), the strain correlation length is 
much larger than the DW coupling.  As $\delta$ is increased to 0.94, the pattern of DW 
domains hardly changes, while the strains are now correlated over essentially the entire field of 
view, Fig.~\ref{fig:2}(d).  For a small additional increase in coupling, $\delta =0.945$, the 
DW domains develop significant correlation, Fig.~\ref{fig:2}(e), and for slightly stronger 
coupling both DWs and strain are fully ordered.  In the stronger coupling regime, Figs.~\ref{fig:2}(e-h), there is
a clear excess of one sign of DW in a given strain domain, an effect which has not been observed in STM\cite{KohsakaPC}.
In contrast, such an excess is weak in the intermediate coupling regime, 
corresponding to Figs.~\ref{fig:2}(a-d), which shows a clear similarity to the STM spectra of 
Ref.~\onlinecite{Seamus}.

While stripes have clearly been seen in LSCO, their presence in other cuprates, including the Bi cuprates and CCOC, is less clear.  Within our model the CDW should be a stable phase.  In earlier STM studies the smectic modulations, or `electronic glass phase', appeared clearly only $\sim$30-40~meV away from the Fermi level.  The modulations were correlated with the pseudogap energy scale, and appeared strongest at energies above 0.6 of the pseudogap energy\cite{JCD}, although they remain visible down to at least 0.2 of the pseudogap, with an energy-independent correlation length\cite{Seamus}. 
At lower energies a similar modulation is present, but lacking contrast reversal.\cite{Hanaguri07,ParkYaz}.  For this feature, 
the first harmonic $\sim$1/4 component is actually found to be strongest near the Fermi level.  In contrast, the third harmonic $\sim$3/4 component is strongest near the pseudogap energy.\cite{JEH}  It may be that superconductivity plays a role in making these low-energy features electron-hole symmetric, an effect missing from our normal state calculation.

While the smectic phase in Bi2212 has not been observed in x-ray diffraction or coherent x-ray scattering studies\cite{Abbano}, this could be an effect of disorder, since the correlation length is about an order of magnitude smaller in Bi2212 than in LSCO.\cite{Wise}  Alternatively, it could be that strain couples the Pomeranchuk mode to a lower-lying mode, and level repulsion prevents the Pomeranchuk wave from going soft.  Interestingly, octahedral tilt and oxygen dimpling modes also couple to strain, and these modes do have low-temperature instabilities in many cuprates\cite{LTT}.  

If the strains are confined to the CuO$_2$ planes, then they would lead
to a local misregistry between Cu and Bi atoms.  A recent STM
experiment has used Zn atoms to probe this, and the results they find
are of the form expected for our model\cite{picome}.  However, these distortions
may be instrumental as they are not reproduced when the direction of the
scan is changed. Alternatively, the strains may couple between planes, with little 
misregistry.  We hope our work will stimulate further experimental
work to determine the real strains.


We have shown that ANN stripes provide a good model for the pseudogap phase in most hole doped cuprates.  
Specifically, the ANN phase in the Bi cuprates has the observed doping dependence of incommensurability, the CO 
displays the contrast reversal seen in STM, and the FS has the non-$(\pi /2,\pi /2)$-centered 
nodal pockets seen in ARPES.  A secondary magnetic order enhances its resemblance to a conventional 
stripe phase.  The phase has quadratic coupling to a $q=0$ shear mode, which may be the anomalous nematic 
phase seen in STM and neutron scattering. The important role of strain in high-T superconducting cuprates has been noted previously.\cite{Bi2}.

{\bf Acknowledgments} This work is supported by the US Department of Energy, Office of
Science, Basic Energy Sciences contract DE-FG02-07ER46352, and
benefited from the allocation of supercomputer time at NERSC,   
Northeastern University's Advanced Scientific Computation Center
(ASCC).  RSM's work has been partially funded by the Marie Curie
Grant PIIF-GA-2008-220790 SOQCS, J.L. is supported by IIT-Seed project NEWDFESCM, 
while GS' work is supported by   
the Vigoni Program 2007-2008 of the Ateneo Italo-Tedesco
Deutsch-Italienisches Hochschulzentrum.  We thank J.E. Hoffman and Y. Kohsaka for many stimulating comments.


\begin{thebibliography}{99}
\bibitem{footV}It is well known\cite{LTT} that shear strain splits the degeneracy of the $x$- and $y$-oriented VHSs, but only if the in-plane oxygens become inequivalent.
\end{thebibliography}
\end{document}


\author{R.S. Markiewicz}\thanks{(Contact: R. Markiewicz: \href{mailto:r.markiewicz@neu.edu}{r.markiewicz@neu.edu})}
\affiliation{Physics Department, Northeastern University, Boston MA 02115, USA}
\author{J. Lorenzana}
\affiliation{ISC-CNR, Dipartimento di Fisica, Universit\`a ``La Sapienza'',
Piazzale Aldo Moro 5, Roma, Italy}
\author{G. Seibold}
\affiliation{Institut F\"ur Physik, BTU Cottbus, PBox 101344, 03013 Cottbus,
Germany}
\author{A. Bansil}
\affiliation{Physics Department, Northeastern University, Boston MA 02115, USA}
\title{
Supplementary Material: Short Range Smectic and Long Range Nematic Order in the Pseudogap Phase of Cuprates}
\maketitle


\subsection{Calculational Details}
Our investigations are based on the following hamiltonian
\begin{equation}\label{eq:mod}
H = H_{e} + H_{el-ph} + H_{ph}
\end{equation}
where $H_{e}$ denotes the Hubbard model, $H_{e-ph}$ the coupling
between electrons and phonons and $H_{ph}$ the bare phonon part.
In the Hubbard model
\begin{displaymath}                                  
H_{e}=\sum_{ij,\sigma}t_{ij} c_{i,\sigma}^\dagger c_{j,\sigma}  
+  U\sum_i n_{i,\uparrow}n_{i,\downarrow}                       
\end{displaymath}                                                  
$c_{i,\sigma}^{(\dagger)}$ destroys (creates) an electron on lattice site $R_i$ and $n_{i,\sigma}=c_{i,\sigma}^\dagger c_{i,\sigma}$.  
We incorporate band structure effects  
by using for the hopping parameters $t_{ij}$ a one-band tight-binding fit 
to the LDA dispersion in Bi-2201.\cite{RM7}  
Interaction effects are incorporated via the (time-dependent) Gutzwiller
approximation [(TD)GA], 
leading to the GA+RPA charge
susceptibility.\cite{DLGS}  

With regard to the electron-phonon coupling $H_{el-ph}$ 
we use  a generic phonon model
consisting of only longitudinal and [in-plane] transverse acoustic branches, 
atomic mass $M$, and electron-phonon coupling $\delta t$.  
The key ingredient is that the phonons modulate the hopping parameter, with 
`longitudinal' meaning the
modulation $\delta t$ varies along the phonon propagation direction, and 
`transverse' meaning
the modulation is at right angles to the propagation.
The corresponding operator reads as
\begin{displaymath}
H_{el-ph}=\sum_{ij,\sigma} t_{ij}\alpha_{ij}  \sum_{\sigma i \mu=x,y} (u^\mu_{j}-u^\mu_i)
(c_{i\sigma}^\dagger c_{j\sigma} + h.c.) 
\end{displaymath}
where $\delta t_{ij}=t_{ij}\alpha_{ij}$ and $u^\mu_{i}$ denotes
the displacement of the atom at site $R_i$ in direction $\mu$.
Finally, the phonon part is given by
\begin{displaymath}
H_{ph}= \frac1{2N}\sum_{\alpha\beta\qvec} u^\alpha_{\qvec} K_{\alpha\beta\qvec} 
u^\beta_{-\qvec}
+ \frac1{2N}\sum_{\alpha\qvec} p^\alpha_{\qvec} \frac{1}{M} p^\alpha_{-\qvec}
\end{displaymath}
which can be diagonalized to yield the
bare phonon frequencies $[\Omega_{q\mu}^0]^2=2(K/M)_{\mu}(2-cos(q_xa)-cos(q_ya))$.
Here $K_{\mu}$, $M_{\mu}$ denote effective spring constant and atomic mass for 
the longitudinal and in-plane transverse ($\mu$ = L [T]) acoustic mode, respectively.

In reality, a strong modulation of
$\delta t$ can be produced by several phonons, including some involving motion of oxygen atoms perpendicular to the CuO$_2$ planes\cite{SeGL}.  Hence in our model, the bare acoustic
frequencies are adjusted to approximate oxygen modes in undoped La$_2$CuO$_4$ [as a generic single-layer cuprate],\cite{WYK} which gives $(K/M)_{LA}=2(K/M)_{TA}\equiv\Omega_L^{02}$, taking $\Omega_L^0=12.4$~meV, and $M$ is the oxygen mass.
In this case, the dominant $q$-vectors, Fig.~\ref{fig:5},  should be 
approximately correct, since
they are controlled by Fermi surface nesting, but the `phase diagram', 
Fig.~\ref{fig:5}, is merely
indicative of regions of strong electron-phonon coupling and significant Kohn 
anomalies.  A realistic calculation of instability would require a model 
of the full phonon dispersion, and
in particular, how the strong softening of one branch interacts with other 
branches of the
same symmetry at lower energy.

The electron-phonon coupling leads to dressed
phonon frequencies according to
\begin{equation}
\Omega_{q\mu\nu}^2=[\Omega_{q\mu}^0]^2\delta_{\mu ,\nu}+\delta K_{\mu ,\nu}/M_{\mu},
\label{eq:26}
\end{equation}
where the renormalized elastic constants can be evaluated within the
TDGA as \cite{oelsen}
\begin{equation}
\delta K_{\mu\nu}=-[\chi^0_{ff\mu\nu}-\tilde\chi^0_{f\mu}\hat
W(1+\tilde\chi_0\hat W)^{-1}\tilde\chi^0_{f\nu}].
\label{eq:27}
\end{equation}

In terms of a bare susceptibility $\chi_{0{\bf q}}=\sum_{\bf k}\chi_{0{\bf k,q}}$, with
\begin{eqnarray}
\chi_{0{\bf k,q}}=
   -{1\over N}\sum_{\sigma}
{n_{{\bf k}+{\bf q},\sigma}-n_{{\bf k},\sigma}\over \epsilon_{{\bf k}+{\bf
q},\sigma}-\epsilon_{{\bf k},\sigma}}.
\label{eq:G3}
\end{eqnarray}
The terms in Eq.~\ref{eq:27} are
\begin{eqnarray}
\tilde\chi_{0{\bf q}}=
  \sum_{\bf k}\chi_{0{\bf k,q}}
\begin{pmatrix}1&E_{{\bf k},{\bf q},\sigma}\\
                               E_{{\bf k},{\bf q},\sigma}&E^2_{{\bf k},{\bf
q},\sigma}\\ \end{pmatrix}
\label{eq:G4}
\end{eqnarray}
\begin{eqnarray}
\tilde\chi^{0}_{f{\bf q}\mu}=
  \sum_{\bf k}\begin{pmatrix}1\\
                               E_{{\bf k},{\bf q},\sigma}\end{pmatrix}
\chi_{0{\bf k,q}}f_{{\bf k}+{\bf q},{\bf k},\mu}^{(1)}
\nonumber\\
+\begin{pmatrix}0\\
              2if_{q,\mu}^{(0)}\\\end{pmatrix},
\label{eq:17}
\end{eqnarray}
\begin{eqnarray}
\chi^0_{ff{\bf q}\mu\nu}=\sum_{\bf k}\chi_{0{\bf k,q}}
f_{{\bf k},{\bf k}+{\bf q},\mu}^{(1)}
f_{{\bf k}+{\bf q},{\bf k},\nu}^{(1)},
\label{eq:19}
\end{eqnarray}
with $ E_{{\bf k},{\bf q},\sigma}= \epsilon^0_{{\bf k}+{\bf q},\sigma}+\epsilon^0_{{\bf
k},\sigma}$, bare dispersion $\epsilon^0_{\bf k}$, dressed dispersion $\epsilon_{\bf
k}=Z\epsilon^0_{\bf k}$, with the Gutzwiller renormalization factor $Z=z_0^2$.
Finally,
\begin{equation}
f_{k,k+q,\mu}^{(1)}=2iz_0^2f_{k,k+q,\mu}^{(0)},
\label{eq:9}
\end{equation}
$f_{k,k+q,\mu}^{(0)}Q_{q}^{\mu}$ is the Fourier transform of the SSH interaction
$f_{i,j,\mu}=t_{i,j}\alpha_{i,j}
(u_j^{\mu}-u_i^{\mu})$, $f_{q,\mu}^{(0)}Q_q^{\mu}$ is the Fourier transform of
$f_{i,\mu}=\sum_jf_{i,j,\mu}$,
$Q_{q}^{\mu}$ is the Fourier transform of $u_i^{\mu}$,
$\alpha_{i,j}=-\partial \ln{(t_{i,j})}/\partial{r}>0$ and $\mu =x,y$.
Equation~\ref{eq:26} can be written in terms of a Stoner factor, which takes a simple form
when $\delta K$ is diagonal [as along $\Gamma\rightarrow (\pi ,0)$, $\Gamma\rightarrow
(\pi ,\pi )$],
\begin{equation}
U_{eff}\chi_{0{\bf q}}=-\delta K_{\mu ,\mu}/M_{\mu}[\Omega_{q\mu}^0]^2.
\label{eq:29}
\end{equation}
It is convenient to normalize $\delta K_{\mu ,\mu}=M_{\mu}E_R^*\delta\hat K_{\mu ,\mu}$, so that the strength of the electron-phonon coupling is measured in terms of
\begin{equation}
E_R^*=\frac{\gamma^2\hbar^2}{Ma^2}=\frac{\lambda_{ep}\hbar^2\Omega_L^{02}}{t},
\label{eq:29b}
\end{equation}
where for nearest neighbor hopping, $\alpha_{i,j}=\gamma /a$ and $a$ is the Cu-Cu separation,
and we have introduced the electron-phonon coupling constant
\begin{equation}
\lambda_{ep}=\frac{\gamma^2t}{Ka^2}.
\label{eq:29c}
\end{equation}
Taking $M$ to be an oxygen mass, $E_{R0}^*\equiv E_R^*/\gamma^2=16.7\mu$eV.  For Bi2201, with $t=435$~meV,
this is equivalent to $\lambda_{ep0}=\lambda_{ep}/\gamma^2=0.047$.  Note that $\lambda_{ep}$ is independent of the phonon mass.

\subsection{Results: Soft phonons, Charge stripes, and Phase Diagrams}

\begin{figure}[H]
\includegraphics[width=15cm,clip=true]{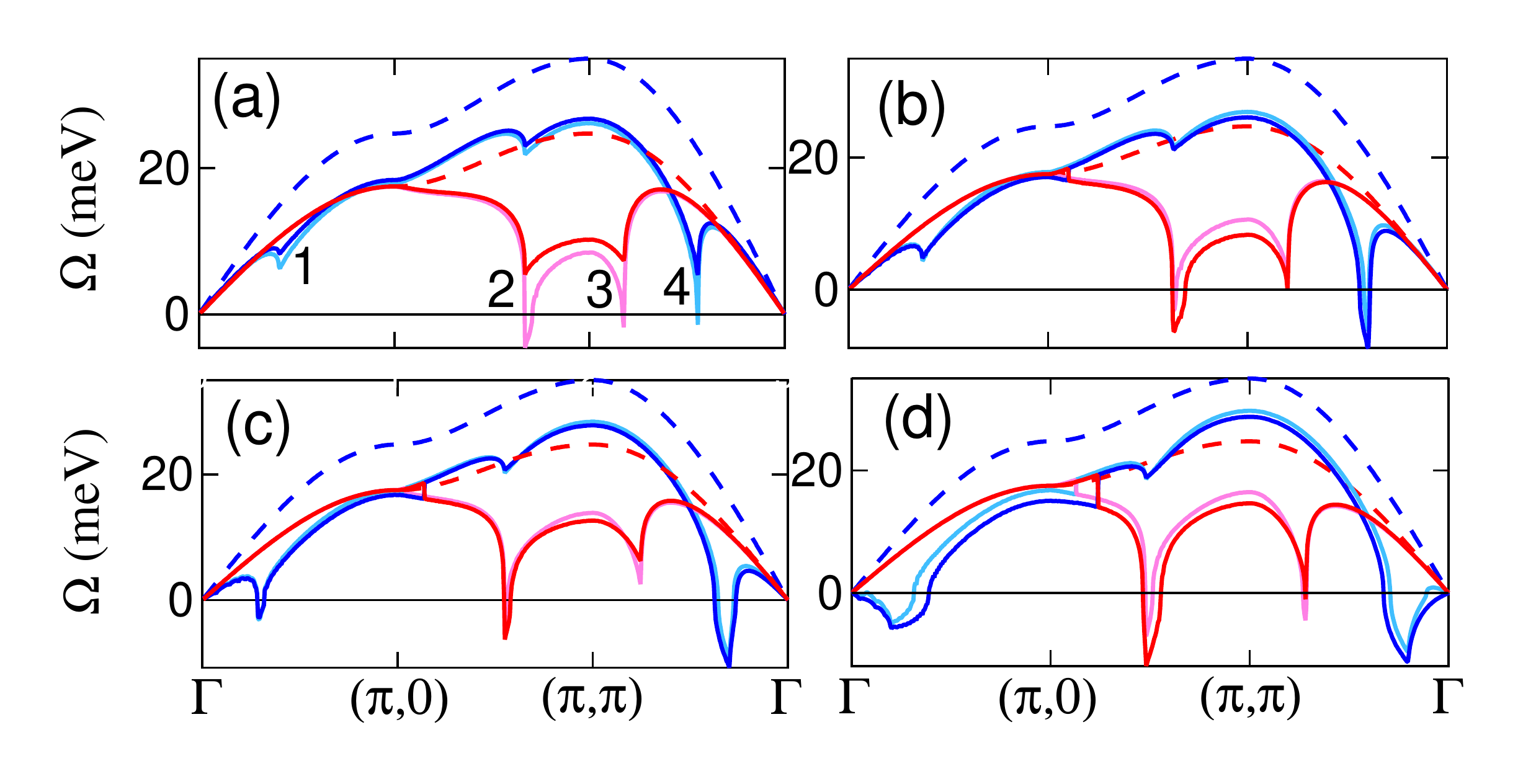}
\caption{
Bare phonon dispersion (dashed line) compared to dressed dispersion assuming $U/U_{BR}$
= 0.20 (light lines) or 0.60 (dark lines) at a series of hole dopings $x$ = (a) 0.05,
(b) 0.10, (c) 0.20, (d) 0.30.  Longitudinal [transverse] phonons in shades of blue
[red].  Material parameters appropriate for Bi-2201, for which $U_{BR}=13.6t$. Only
modulation of the nearest-neighbor hopping $t$ is included, with doping independent
magnitude $\gamma =3.28$.
By convention, real $\Omega_{ph}$'s are
plotted as positive numbers, imaginary $\Omega_{ph}$'s as negative.
}
\label{fig:5}
\end{figure}

Figure~\ref{fig:5} compares the bare and renormalized LA and [in-plane] TA phonon
frequencies in Bi2201.  Along the $(\pi ,0)\rightarrow (\pi ,\pi )$-branch the modes are
mixed, and labelled as predominantly longitudinal or transverse.  The sharp dips in the dressed frequencies are
caused by peaks in the bare susceptibility associated with FS nesting.  Each peak in $\chi$
leads to a prominent Kohn anomaly in the phonon spectrum, which can lead to an instability
if the renormalized $\Omega_{ph}^2$ becomes negative. By comparing the present results with
earlier calculations for magnetic stripes,\cite{RM7} we find that the instabilities fall at
nearly the same $q$-values for both kinds of stripes as a function of doping, being
controlled by the same Fermi surface nesting.

In the doping range appropriate to the cuprates, there are two
competing instabilities.  For very low doping the dominant instability is to an
incommensurate state at the edge of the $(\pi ,\pi )$ plateau at $(\pi -\delta ,\pi
)$ [anomaly 2 in Fig.~\ref{fig:5}(a) -- called a {\it plateau} instability], while at higher
dopings a diagonal ANN phase is dominant at $(\pi -\delta ,\pi -\delta )$ [4 in
Fig.~\ref{fig:5}(a)].  Figure~\ref{fig:5} compares the phonon softening in a weak coupling
($U/U_{BR}=0.2$) and an intermediate coupling case ($U/U_{BR}=0.6$) more appropriate
for cuprates.

\begin{figure}[H]
\includegraphics[width=12cm,clip=true]{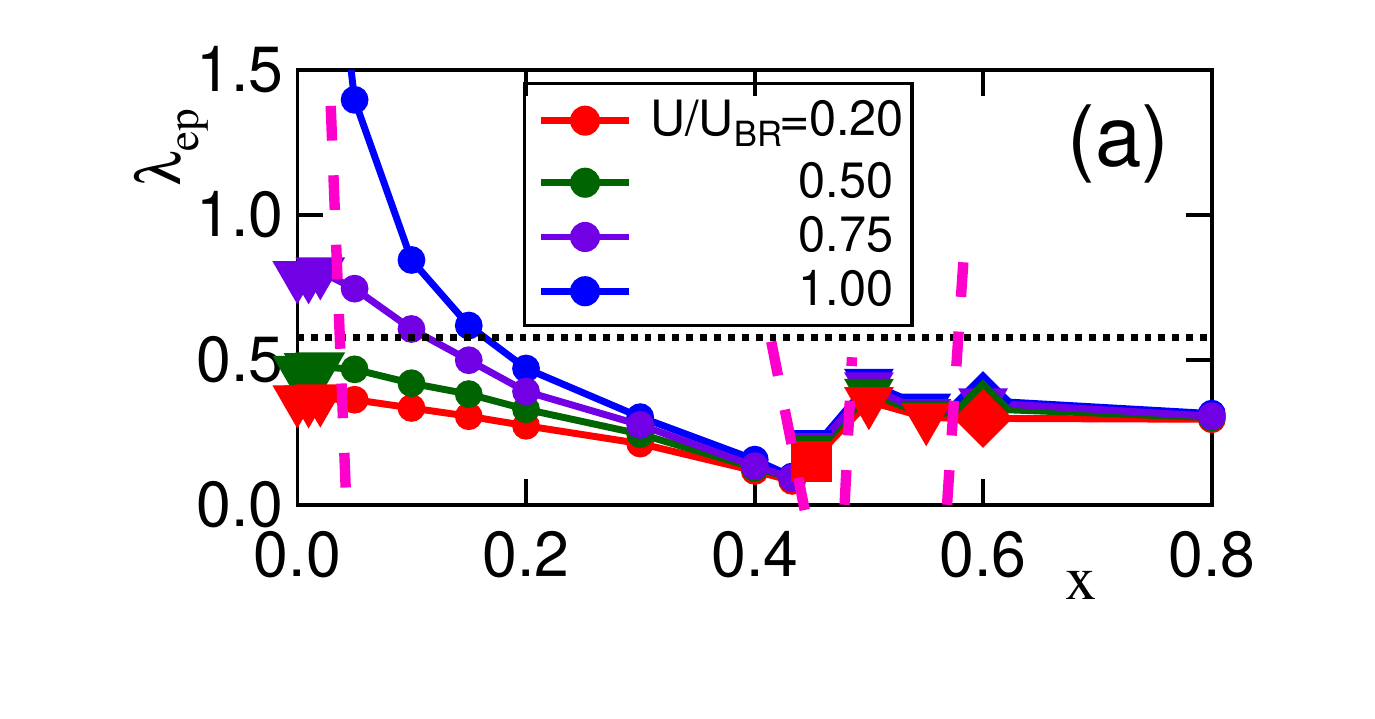}
\includegraphics[width=12cm,clip=true]{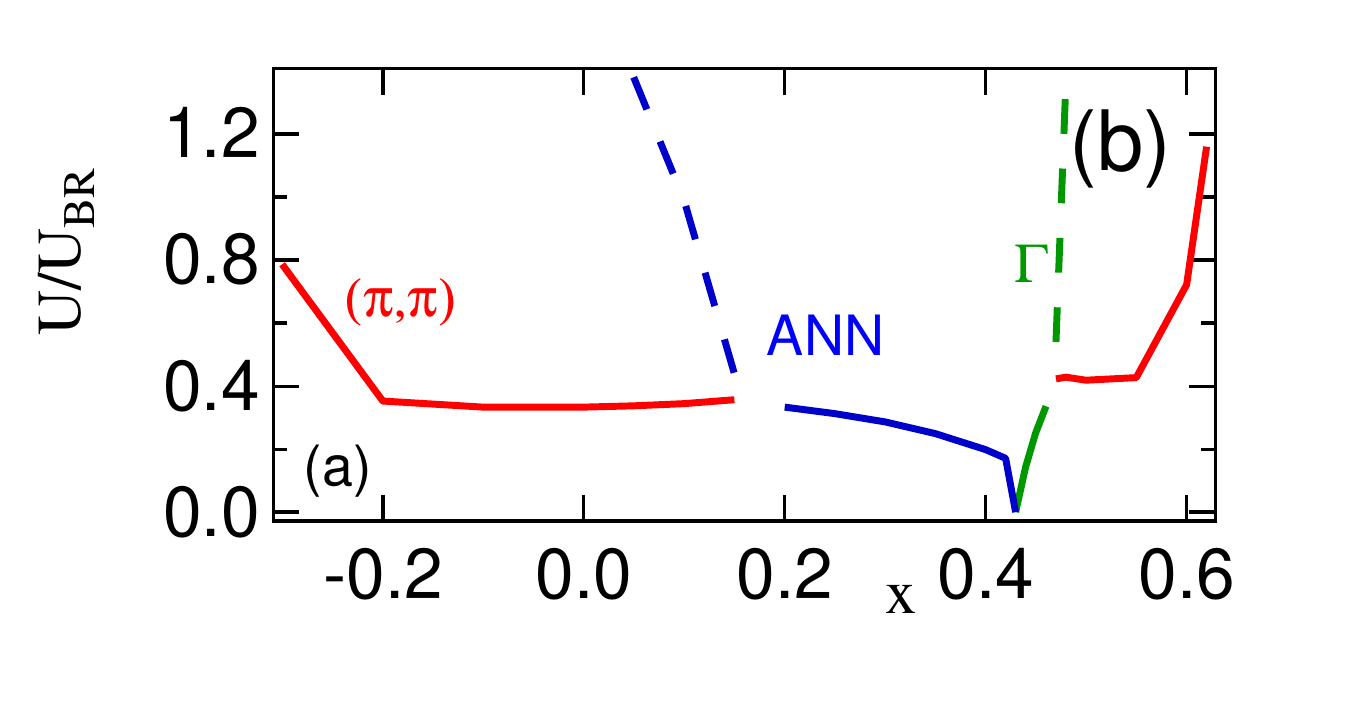}
\caption{
{\bf Phase diagrams for Bi2201.} (a) Charge order, showing threshold strength of electron-phonon
coupling $\lambda_{ep}$ for several values of Hubbard $U$.  Symbols represent different stripe
symmetries, as vertical [diagonal] $(\pi ,\pi )$-plateau stripes = triangles [diamonds]; vertical
[diagonal] ANN stripes = squares [circles].  Dashed lines indicate transitions between different symmetries, while dotted line corresponds to $\gamma =3.5$.  (b) Magnetic order phase diagram, from Ref.~\onlinecite{RM7}.
}
\label{fig:6}
\end{figure}

Figure~\ref{fig:6}(a) shows the resulting charge order phase diagram for several values of
$U$.  As expected, when $U$ is large ($\ge U_{BR}$, the Brinkman-Rice coupling) charge
order is strongly suppressed at half filling, but it is restored rather quickly when $U$ is
reduced or $x$ increased.  ANN stripes (circles and squares) are dominant for a wider doping range than in the magnetic phase diagram, but near half filling and in the overdoped regime vertical/horizontal
stripes associated with the $(\pi ,\pi )$-plateau (triangles and diamonds) win out.  Near
the VHS ($x=0.42$) the value of the critical parameter goes to zero, and it is found that
the dominant nesting $q$-vector approaches $\Gamma$, corresponding to phase separation.
There is an overall similarity with the magnetic-order phase diagram, Fig.~\ref{fig:6}(b),
found earlier.\cite{RM7}  Similar phase diagrams have been found for most cuprates studied,
except for the magnetic phase diagram of LSCO, for which the VHS is much closer to half filling, resulting in the
dominance of the $(\pi ,\pi )$-plateau instabilities.

Figures~1 and~\ref{fig:6} give a very graphic picture of the role of Fermi
surface nesting in the cuprates: not only are the dominant $q$-vectors correctly predicted,
but also the correct phase between two competing DW instabilities.  However, exact
agreement between the FS nesting vector and the experimental DW vector is not to be
expected, even for conventional CDWs and SDWs.  First, nesting is a manifestation of a
Stoner-type instability, $1-U\chi_0({\bf q},\omega =0) =0$, where $U$ is here a
Gutzwiller-modified Hubbard $U$ and $\chi_0$ is a bare [charge or magnetic] susceptibility.
Only at threshold $U=U_c$ is FS nesting exact.  For $U>U_c$ the dominant $q$-vector adjusts
to further lower the electronic energy.  Numerical estimates suggest the shift is relatively
small unless the FS is fully gapped.$^{12}$  One common effect in conventional DWs is a
commensurability pinning: the DW can gain energy by shifting from an incommensurate nesting
vector to a nearby commensurate vector.\cite{McMillan}  This is likely to be
important in the cuprates, where the DW periodicities correspond to only a few atoms across.
For instance, several cuprates seem to have the DWs pinned near a 4-Cu
periodicity, and the Bi-2201 data in Fig.~1 resemble a crossover from period 4
to period 6 [near nesting] with increasing doping.  For Bi-2212, the $q$-vector is more consistent with the prediction for the bonding band.  It may be that proximity to period-4 order stabilizes the bonding band DW over the competing antibonding band DW.  Another factor is that in two-dimensions
the lowest-energy DW state corresponds to double nesting -- simultaneously nesting two
disconnected FS sections.  This can be achieved either by a single DW or by two or more
coherent DWs\cite{McMillan}.  As noted above, the latter seems to be the case for the ANN
phase in the Bi-cuprates: the spatial pattern seen in Bi-2201$^{6}$ can be fit as a
coherent superposition of horizontal and vertical stripes.  The superposition of orthogonal CDWs forms a distinct pattern from a checkerboard.  See Fig.~1(c) of Ref.~[6].
Finally, even when the DW represents a local minimum of the free energy, the
global minimum may correspond to a state of [nanoscale] phase separation, as 
predicted in Bi-2212.$^{12}$

\subsection{Period 6 Stripes}

\begin{figure}[H]
\includegraphics[width=15cm,clip=true]{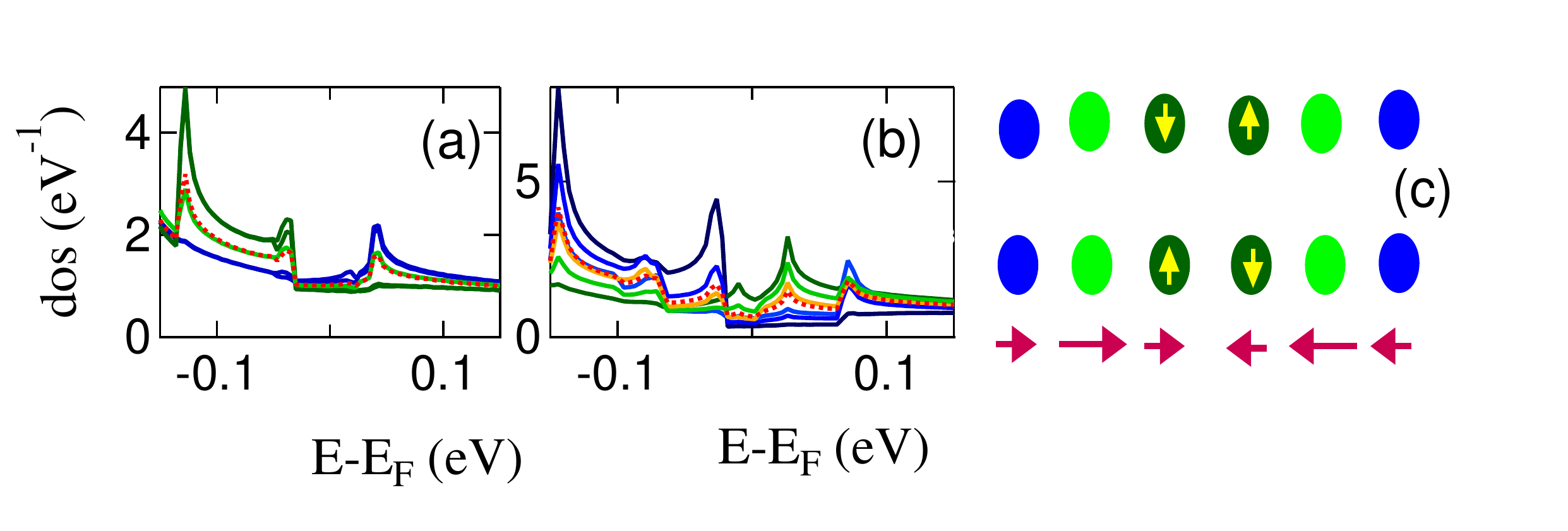}
\includegraphics[width=15cm,clip=true]{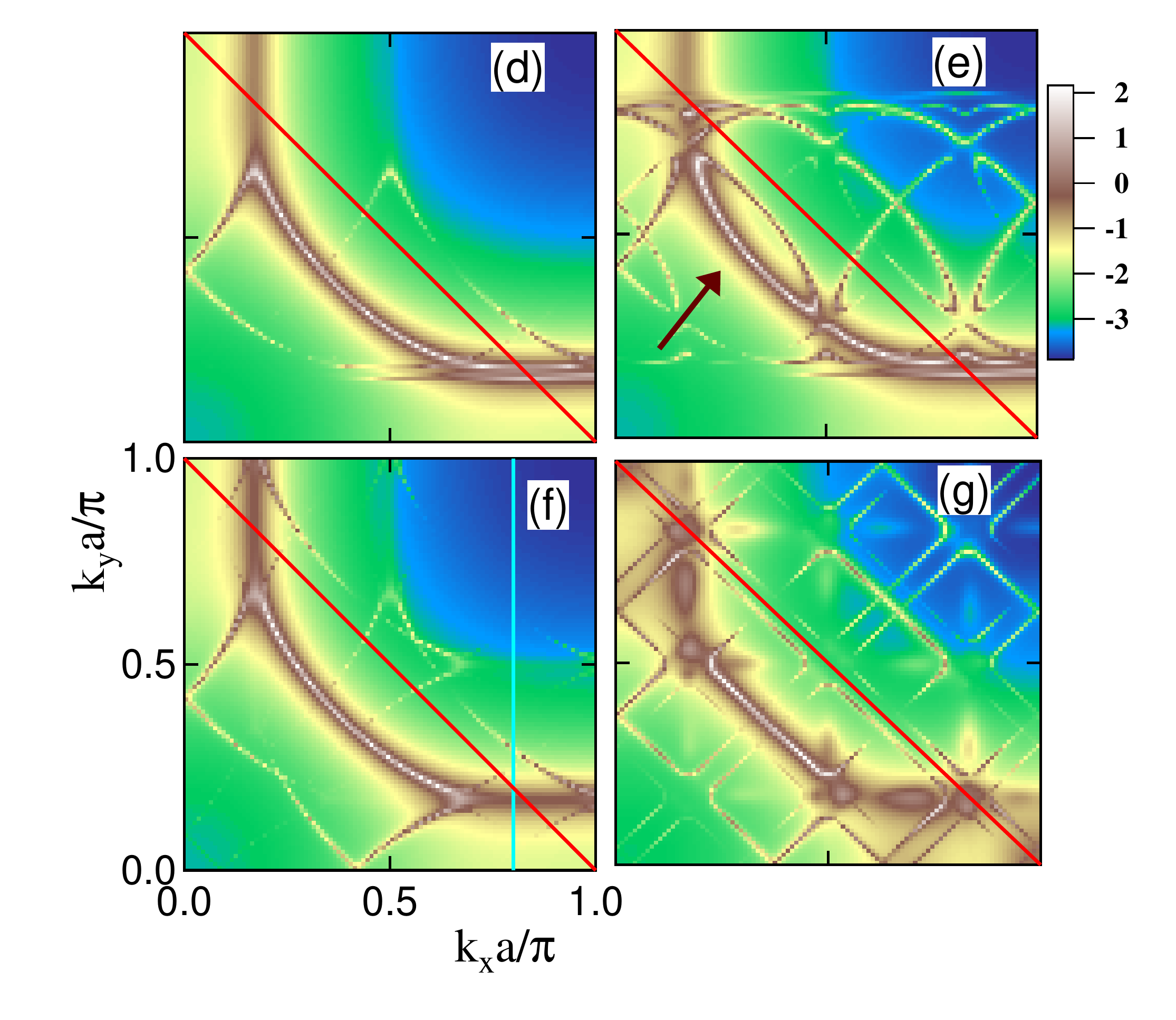}
\caption{
ANN superlattice for Bi-2212. (a,b) Average (red) and local dos for 1D (a) and 2D (b)
stripes.  Defining $s$ and $l$ for short and long bonds, a site with two short bonds as
$(s,s),$ etc., the various local dos are represented as follows: in (a): $(s,s)$ (blue),
$(s,l)$ (light green), and $(l,l)$ (green).  In (b): $(s,s)_x, (s,s)_y$ (dark blue),
$(s,s)_x, (s,l)_y$ (blue), $(s,s)_x, (l,l)_y$ (light blue), $(s,l)_x, (s,l)_y$ (orange),
$(s,l)_x, (l,l)_y$ (light green), and $(l,l)_x, (l,l)_y$ (green).
(c) Local distortion pattern for 1D stripes.  Also shown is pattern of AFM order assumed
for Figs.~S3(f,g).  (d,f) Fermi surface maps for 1D (d) and 2D (f) stripes, with spectral weight plotted on a logarithmic scale.  (e,g) Corresponding FS maps with added AFM order.
Arrow in (e) illustrates a prominent half-pocket Fermi surface section.
}
\label{fig:3}
\end{figure}

Further insight into the ANN phase comes from comparisons with photoemission and tunneling data.  This requires real space modeling, as in Fig.~\ref{fig:3}.
We adjust the doping to give a simple commensurate superlattice vector $Q_0$. For
definiteness we model Bi2212, neglecting bilayer splitting (dotted line in Fig.~1).  In this case $Q_0$ is exactly $2\pi /6a$, where $a$ is the lattice constant, at a doping
$x=0.19$.  We model the one-dimensional ANN order as a Pomeranchuk wave, with modulated
hopping parameter $t_x= t\pm \delta t$, $t_y=t$ with $\delta t$ proportional to the lattice
distortion and the maximum $\delta t=0.1t$, neglecting possible
modulations of $t'$, $t''$, and $t'''$.  We assume a bond-centered lattice distortion,
Fig.~\ref{fig:3}(c).  Since the STM images look more like 1D stripes in Bi2212, and crossed (2D) stripes in Bi2201, we model both, with the 1D results in Fig.~\ref{fig:3}(a,d,e) and the
2D results in Fig.~\ref{fig:3}(b,f,g).  For this distortion, Figs.~\ref{fig:3}(a,b) show the
density-of-states (DOS), both the average value (red lines) and the local values on
different sites.  There is a local pseudogap on the stripes, of a magnitude
comparable to that seen in the STM experiment.  For the 1D stripes, Fig.~\ref{fig:3}(a) shows
that the short-bond rows have a larger local DOS for filled states.  There is a
clear {\it contrast reversal} between electron and hole states -- a higher DOS for
filled [empty] states on the large-t [small-t] atoms. Similar results arise
for the 2D patterns, Fig.~\ref{fig:3}(b). An approximate contrast reversal is commonly found in CDW systems\cite{CDWcr},
and the resulting pattern bears a striking resemblance to the phase seen in STM experiments.$^{4,5,13,14}$  Figures~\ref{fig:3}(d,f) show the resulting FSs, with a structure
factor (SF) correction appropriate for angle resolved photoemission (ARPES)
spectra\cite{MKstripe,MKstripe2}.  The spectral weight is plotted on logarithmic scale to enhance
weak superlattice features.
When the SF is included, it is seen that most of the spectral weight lies near the original
FS.  For either 1D or 2D stripes, the FSs are composed of nodal arcs, as seen in
experiment, with weaker features near the antinodes.

The present calculation represents a minimal model of the stripes.  We note that a
number of refinements are possible.  First, in principle one could also include an
on-site (Madelung) correction of $\sim$+[-]50~meV per short [long] bond, as holes are
attracted by the negatively charged oxygens.  In the absence of this term, we find that
holes tend to pile up on the atoms with more long bonds, and avoid the atoms with short
bonds.  Moreover, a number of {\it secondary order parameters} are not only possible, but
likely, greatly complicating the DW picture.  As can be seen from the FS maps,
Figs.~\ref{fig:3}(d,f), there are a number of level crossings that the Pomeranchuk wave has not gapped.  Additional electronic distortions can take advantage of this, opening gaps at
the crossing points to lower their energy and further stabilize the distortion.

Here we give a single illustration of this effect.  While this is a conventional charge
density wave (CDW), in the cuprates it can acquire aspects associated with stripe physics.
Thus, rows with lower doping will display enhanced Mott physics while the hole-doped rows
would be closer to optimal doping.  In Fig.~\ref{fig:3}(c) we show an example where proximity to half
filling induces a secondary AFM order on the less doped stripes.  Fig.~\ref{fig:3}(e) shows the resulting FS map in the 1D case.  Remarkably, the Fermi arcs have become pockets (arrow), which are not centered on $(\pi
/2,\pi /2)$, as observed in some experiments.$^{17-20}$ Similar effects
arise in 2D, Fig.~\ref{fig:3}(g), but here optimal doping shifts to higher $x=0.31$ to better nest the
antinodal regions.  Note that the Fermi arc seems to terminate along the AF zone boundary,
consistent with STM$^{5}$, while the FS contour is quite squarish, which has been taken
as a signature of underlying stripe order in ARPES studies.$^{15}$


Recent quantum oscillation (QO) studies have found evidence for small FS pockets in both
hole-doped YBa$_2$Cu$_3$O$_{7-\delta}$ (YBCO)\cite{QOsc,QOsc3,QOsc4} and electron-doped Nd$_{2-x}$Ce$_x$CuO$_4$ (NCCO)\cite{QOHelm}.
While the present model successfully predicted the areas in NCCO\cite{Kusko}, earlier proposed FSs\cite{QOsc2,QOsc5} for
YBCO bore little resemblance to the present results.  However, after this manuscript was completed,
we became aware of new results which found that there is most
likely only a single (electron-like) pocket in underdoped YBCO\cite{Seba2}, which can be modeled by a two-$q$ density
wave\cite{Seba3}, similar to the present result.  While the doping range in YBCO where these QOs
are observed probably corresponds to period-4 stripes, the proposed Fermi surface preserves the
nodal states, and overall bears a close resemblance to the high spectral weight star-shaped feature
in Fig.~S3f.  Indeed, we find that a similar nodal Fermi surface exists in Bi-2212 in this
period-6 crossed stripe state.

\begin{figure}
\includegraphics[width=9cm,clip=true]{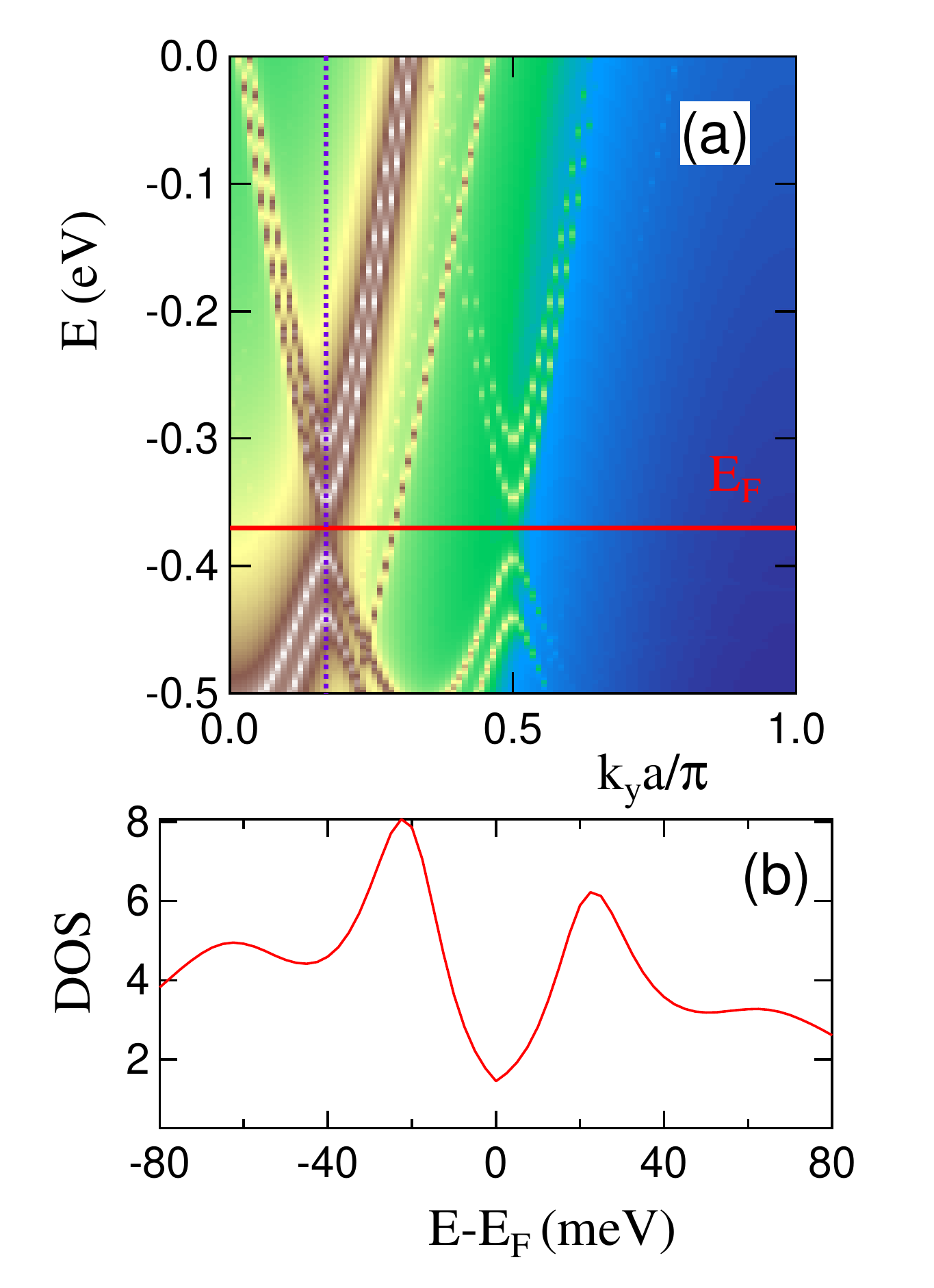}
\caption{
(a) $E-k$ map of electronic dispersion along the cut in Fig.~S3(f) [blue line], showing ANN gap centered at the Fermi level $E_F$.  (b) Plot of spectral weight vs binding energy along the dotted line in frame (a).  For (b), an energy dependent broadening $\gamma =0.33\omega +5$~meV was included to mimic the experimental results.
}
\label{fig:4}
\end{figure}

Finally, we note that ANN nesting provides a natural explanation for one more feature of the ARPES data.  Yang, et al.$^{17}$ noted that the pseudogap in Bi2212 is {\it centered at the Fermi level} in the antinodal regime.  As they point out, this is unexpected for a simple $(\pi ,\pi )$ antiferromagnetic order.  However, since the ANN phase specifically involves nesting along the long flat sections of the FS in the antinodal regime, it is natural to expect that this nesting is centered at the Fermi level.  We show that this is in fact the case in Fig.~\ref{fig:4}.  Figure~\ref{fig:4}(a) shows the dispersion along a cut in the antinodal regime, corresponding to the light blue line in Fig.~\ref{fig:3}(f).  It can be seen that the ANN gap is centered on the Fermi level.  Further confirmation is shown in Fig.~\ref{fig:4}(b), which plots the spectral weight along the dotted line in Fig.~\ref{fig:4}(a).  To compare with experiment [Fig.~4(f) of Ref.~[17]], the data in this figure have been broadened by $\gamma =0.33\omega +5$~meV.  Clearly the ANN nesting provides a natural (non-superconducting) explanation for
this feature.

Half pockets, as in Fig.~\ref{fig:3}(e), are also found in a phenomenological model of the cuprates based on the $t-J$ model.\cite{YRZ}  However, this model cannot explain CO with nesting vectors different from  $(\pi ,\pi )$, and must explain the pseudogap centered at the Fermi level as a consequence of a superconducting gap.$^{17}$  But this is inconsistent with growing evidence that superconductivity only appears at an onset temperature $T_{onset}<T^*$, the pseudogap temperature.\cite{twopgs,twopgs2,twopgs3,twopgs4,twopgs5,twopgs6,twopgs7}

The present model bears some resemblance to the valence bond glass
(VBG)\cite{VBG2,ValenceBondGlass}.  Indeed, since impurities are a relevant perturbation of a CO phase, the ANN phase will become more glass-like in a real Bi-2212 sample with large interstitial-oxygen disorder.  However, a significant difference is that the one-dimensional ANN phase has no bond length modulation along $y$ (the direction along the stripe), while the VBG has a $d$-wave-like modulation.

\section{Summary of Model Calculations}

The present letter plus supplementary material describes three closely related calculations which together provide a coherent microscopic model for charged stripes, to complement our earlier work on magnetic stripes.\cite{RM7}  The three calculations are designed to answer three questions: (1) what is the phase diagram for predominantly charge-based stripes?
(2) can these stripes explain the various features of stripes seen in ARPES, STM, and quantum oscillation measurements? and (3) how are the `nematic' phenomena related to these stripes?  Each calculation requires progressively more approximations, so in this subsection we summarise them.

The underlying microscopic model contains three key ingredients.  (1) The dominant electron-phonon interaction arises from modulations of Cu-Cu bondlength, which modify the electronic hopping energies.  (2) Since the energies change as high powers of the bondlengths, there is also a second-order ($t^2$) coupling to the electronic nematic order, which couples directly to elastic shear strain.  (3) The strain coupling is long range, so that even if charge pinning restricts the stripes to nanoscopic length scales, the strain-nematic order can persist over substantially larger length scales, as found in experiment.  We have amply documented all three ingredients, so that the model qualitatively describes the observed phenomena.  The question remains, to what extent is the agreement quantitative?

For the {\it phase diagram} calculation [Figs.~1, S1, and~S2], the electronic parameters [hopping energies and Hubbard $U$] are the same as used for our magnetic calculations, and should be quite accurate.  The electron-phonon interaction parameter $\gamma$ is consistent with literature values, although it is not clear how the $\gamma$s vary for more distant hoppings, so we have simply neglected higher order terms.  The main assumption is in the bare phonon frequency $\Omega_0(q)$.  We have introduced an effective one-band model, to approximately describe the Cu displacements and corresponding in-plane `acoustic' modes.  A more realistic model would be desirable, but would require a model of the  full `bare' phonon dispersion, including eigenfunctions for all branches, in the absence of this electron-phonon coupling term.  We believe that the present model properly describes the competition between Hubbard-$U$ suppression of charge order and the tendency of electron-phonon coupling to form stripes, and captures the dominant Fermi-surface nesting vectors.  The approximations on the phonon spectrum will mainly affect the exact coupling $\gamma$ for instability and the nature of the phonon mode that condenses.  For example, in LSCO the low-temperature orthorhombic and low-temperature tetragonal phases both involve soft modes of octahedral tilts.  Interestingly, both modes couple to shear strains, and both do go unstable in appropriate doping ranges. 

The phase diagram gives the minimum $\gamma$ (or equivalently, electron-phonon coupling $\lambda$) needed to produce a CDW order, and the corresponding $q$-vector.  To find the corresponding {\it real and $k$-space stripe structure} [Figs.~S3,4] we should choose a $\gamma$ larger than the threshold value and numerically minimize the free energy on a lattice, which will produce a shifted $q$-vector, as found in the magnetic stripe case, Ref.~[16].  Instead, we simply assumed a plausible amplitude for the lattice distortion and calculated the corresponding electron energy levels.  We also estimated the effect of a secondary magnetic order, as expected for stripes near half filling.  We believe that such approximations are adequate to capture the qualitative features of the experiments.

Finally, the {\it nematic order} arises naturally in the presence of a shear strain which splits the VHS degeneracy.  We demonstrated how a shear strain arises as a secondary order parameter in the presence of stripes.  The {\it emergent} nature of the nematic phase arises from the long-range nature of the strain coupling, which makes it much more robust than the stripe order in the presence of point impurity disorder.  In calculating Fig.~2 we were interested in one issue: could the nematic order have a much larger correlation length that the underlying stripe order, in the presence of impurity pinning?  In the main text we have provided a concrete calculation showing that the answer is yes, while above we presented a possible microscopic realization.

[H]
